 \documentclass[final,5p,times,twocolumn,authoryear]{elsarticle}

\usepackage{amssymb}
\usepackage{lipsum}
\usepackage{amsmath}
\usepackage{booktabs}
\usepackage[colorlinks=true
  ,urlcolor=blue
  ,anchorcolor=blue
  ,citecolor=blue
  ,filecolor=blue
  ,linkcolor=blue
  ,menucolor=blue
  ,linktocpage=true
  ,pdfproducer=medialab
  ,pdfa=true
]{hyperref}
\usepackage{orcidlink}

\journal{Physics Letters B}

\begin{document}

\begin{frontmatter}



\title{White Dwarf Structure in $f(Q)$ Gravity}


\author{Rajasmita Sahoo\,\orcidlink{0000-0002-9265-5025}}
\affiliation{organization={National Institute of Technology},
            city={Tiruchirappalli}, 
            state={Tamil Nadu},
            postcode={620015},
            country={India}}

\begin{abstract}
In this work, we investigate the equilibrium structure of white dwarfs within the covariant formulation of symmetric teleparallel $f(Q)$ gravity, in which gravity is described by the nonmetricity scalar $Q$ instead of spacetime curvature. We consider static and spherically symmetric stellar configurations composed of cold, fully degenerate electron matter and adopt a quadratic form of the gravitational Lagrangian, $f(Q)=Q+\alpha Q^{2}$, where $\alpha$ quantifies deviations from general relativity. The corresponding modified stellar structure equations are solved numerically in conjunction with the Chandrasekhar equation of state.
We examine the impact of the parameter $\alpha$ on the internal structure and global properties of white dwarfs, including the radial profiles of the metric potentials, pressure, density, nonmetricity scalar, and enclosed mass, as well as the mass--radius relation. While negative values of $\alpha$ were explored, they lead to unstable or nonphysical configurations at high densities; therefore, the analysis is restricted to non-negative values of $\alpha$.
Our results show that nonmetricity corrections produce significant deviations from the general relativistic predictions in the high-density regime. In particular, increasing $\alpha$ modifies the equilibrium configurations and leads to a reduction in the maximum mass relative to the Chandrasekhar limit, accompanied by corresponding changes in the stellar radius and interior profiles. For $\alpha = 5\times10^{18}\,\mathrm{cm^2}$, we obtain a maximum mass $M_{\max}=1.3519\,M_{\odot}$ and radius $R=2228.85\,\mathrm{km}$, which are consistent with the observational constraints of the ultra-massive white dwarf ZTF J1901+1458. 
These findings suggest that white dwarfs can provide a complementary astrophysical probe for testing the viability of $f(Q)$ gravity in the strong-field regime.
\end{abstract}

\begin{keyword}
White dwarfs \sep Modified gravity \sep f(Q) gravity  \sep Mass-radius relation
\end{keyword}
\end{frontmatter}
\section{Introduction}
General Relativity (GR) has been remarkably successful in describing gravitational phenomena across a wide range of length scales. Nevertheless, several observational discoveries in modern cosmology, particularly the accelerated expansion of the Universe inferred from Type Ia supernova observations and other cosmological probes, suggest that GR may require extensions or modifications in certain regimes~\cite{SupernovaSearchTeam:1998fmf,SupernovaCosmologyProject:1998vns,Planck:2015fie,Planck:2015bpv}. In addition to cosmic acceleration, phenomena such as galaxy rotation curves, gravitational lensing anomalies, and the formation of large-scale structures~\cite{rubin70,clowe06} are often interpreted through the introduction of dark matter and dark energy. These challenges have motivated extensive efforts to explore modified gravity theories (MGTs), which aim to extend or generalize GR in order to achieve a deeper understanding of gravitational interactions at cosmological and astrophysical scales.
In the standard formulation of GR, gravity is described within the framework of Riemannian geometry, where the affine connection is metric compatible and torsion-free, corresponding to the Levi--Civita connection. In this geometric description, spacetime curvature is responsible for mediating gravitational interaction. However, alternative formulations of gravity can be constructed by relaxing these geometric assumptions~\cite{BeltranJimenez:2019esp,Harada:2020ikm}. Besides curvature, spacetime geometry may also include torsion and nonmetricity as fundamental geometrical quantities. For instance, by choosing a connection with vanishing curvature and nonmetricity but non-zero torsion, one obtains the teleparallel equivalent of general relativity (TEGR), where gravity is described by the torsion scalar $T$~\cite{Aldrovandi:2013wha,Maluf:2013gaa}. Alternatively, if both curvature and torsion vanish while nonmetricity remains non-zero, gravity can be formulated within the framework of symmetric teleparallel gravity (STG), where gravitational interaction is mediated by the nonmetricity scalar $Q$~\cite{Nester:1998mp,Adak:2005cd,Mol:2014ooa,Jarv:2018bgs,BeltranJimenez:2017tkd,Gakis:2019rdd}.

A natural extension of symmetric teleparallel gravity is obtained by generalizing the gravitational Lagrangian from the linear scalar $Q$ to an arbitrary function $f(Q)$, leading to the socalled $f(Q)$ gravity theory~\cite{BeltranJimenez:2019tme}. In this framework, the dynamics of gravity are governed by nonlinear functions of the nonmetricity scalar, allowing for richer gravitational behavior compared with GR. In recent years, $f(Q)$ gravity has attracted considerable attention, particularly in cosmological contexts where it has been explored as a possible explanation for the accelerated expansion of the Universe and other large-scale cosmic phenomena~\cite{Lazkoz:2019sjl,Frusciante:2021sio,Atayde:2021pgb,Lu:2019hra,BeltranJimenez:2019tme,Barros:2020bgg,Frusciante:2021sio,Anagnostopoulos:2021ydo,Khyllep:2021pcu,Narawade:2022jeg,Narawade:2022cgb,Heisenberg:2023lru,Nojiri:2024zab}. Several cosmological models constructed within this framework have shown that modified nonmetricity theories can provide viable alternatives to the standard $\Lambda$CDM model. Beyond cosmology, $f(Q)$ gravity has also been applied to compact astrophysical systems, including neutron stars, hybrid stars, and quark stars, where the strong gravitational fields and extreme matter densities provide an excellent arena for testing deviations from GR~\cite{Lin:2021uqa,Alwan:2024lng,Sharif:2025rgz,Sokoliuk:2022bwi,Harko:2018gxr}.

In most studies of $f(Q)$ gravity, the theory is formulated using the coincident gauge, in which the affine connection can be chosen to vanish identically through an appropriate coordinate transformation~\cite{BeltranJimenez:2017tkd,BeltranJimenez:2019esp}. In this gauge, the covariant derivative reduces to the ordinary partial derivative and the nonmetricity tensor simplifies to $Q_{\alpha\beta\gamma}=\partial_{\alpha}g_{\beta\gamma}$, which considerably simplifies the field equations. While this formulation has been widely used in cosmological applications, it relies on a specific coordinate choice and may obscure the underlying geometric structure of the theory.
To overcome this limitation, a fully covariant formulation of $f(Q)$ gravity has been developed, where the affine connection is treated as an independent geometric object and is not fixed to vanish~\cite{BeltranJimenez:2018vdo,Harko:2018gxr}. In this approach the gravitational dynamics are described in a coordinate--independent manner, allowing a more general treatment of spacetime geometry. The covariant formulation has been successfully applied in several contexts, including cosmology and the study of compact astrophysical objects, providing a consistent framework for analyzing gravitational systems beyond the coincident gauge~\cite{Dialektopoulos:2025ihc,Alwan:2024lng}.

Among compact stellar remnants, white dwarfs represent the final evolutionary stage of the majority of stars and play a fundamental role in stellar astrophysics. The equilibrium structure of a white dwarf arises from the balance between gravitational collapse and the pressure of a degenerate electron gas, leading to the well-known Chandrasekhar mass limit within GR~\cite{Chandrasekhar1931,chandra35,shapiro83}. Typical central densities of white dwarfs can reach values of order $10^{9}$--$10^{10}\,\mathrm{g\,cm^{-3}}$, making them particularly suitable systems for probing gravitational physics in high-density environments.

Recent astrophysical observations have further strengthened the motivation to study white dwarfs within modified gravity frameworks. Observations of peculiar and overluminous Type Ia supernovae suggest the possible existence of super-Chandrasekhar white dwarfs with masses exceeding the classical Chandrasekhar limit~\cite{SNLS:2006ics,scalzo2010nearby}. Conversely, several studies indicate the presence of sub-Chandrasekhar progenitors associated with underluminous Type Ia supernovae~\cite{woosley2011sub,Sim:2010kc}. In addition, high-precision measurements of white dwarf masses and radii obtained from modern astronomical surveys and missions such as \textit{Gaia} have provided increasingly accurate constraints on the white dwarf mass--radius relation~\cite{kilic2018gaia,joyce2018testing}. These observational developments highlight the importance of investigating white dwarf configurations within alternative theories of gravity.
Motivated by these considerations, in this work we investigate the structure of white dwarfs in the framework of covariant $f(Q)$ gravity. In particular, we consider the quadratic model $f(Q)=Q+\alpha Q^{2}$ and derive the corresponding modified Tolman--Oppenheimer--Volkoff (TOV) equations governing stellar equilibrium. Using the equation of state for a degenerate electron gas, we compute the internal structure and mass--radius relations of white dwarfs and examine how the parameter $\alpha$ influences their global properties. Our analysis provides new insights into the role of nonmetricity based modifications of gravity in compact stellar systems and offers a potential avenue for testing alternative gravitational theories through white dwarf observations.

The paper is organized as follows. In Sec.~\ref{sec:2} we briefly review the geometric formalism of covariant $f(Q)$ gravity and present the corresponding field equations. In Sec.~\ref{sec:3} we derive the modified stellar structure equations for static and spherically symmetric configurations in covariant $f(Q)$ gravity. In Sec.~\ref{sec:4} we describe the equation of state used to model the degenerate electron gas in white dwarfs. The numerical results and their physical implications are presented in Sec.~\ref{sec:5}. Finally, in Sec.~\ref{sec:6} we summarize our main conclusions.
\section{Geometric framework of covariant $f(Q)$ gravity}
\label{sec:2}

In symmetric teleparallel gravity, the gravitational interaction is described by the nonmetricity of spacetime rather than curvature or torsion. Within this framework, the affine connection can be decomposed into three independent contributions,
\begin{eqnarray}
\Gamma^{\alpha}{}_{\beta\gamma}
=\left\{^{\alpha}{}_{\beta\gamma}\right\}
+K^{\alpha}{}_{\beta\gamma}
+L^{\alpha}{}_{\beta\gamma},
\label{eq:1}
\end{eqnarray}
where $\left\{^{\alpha}{}_{\beta\gamma}\right\}$ denotes the Levi--Civita connection constructed from the metric tensor $g_{\beta\gamma}$, $K^{\alpha}{}_{\beta\gamma}$ is the contortion tensor associated with torsion, and $L^{\alpha}{}_{\beta\gamma}$ is the disformation tensor arising from nonmetricity. The Levi--Civita connection is defined as
\begin{eqnarray}
\left\{^{\alpha}{}_{\beta\gamma}\right\}
=
\frac{1}{2} g^{\alpha\lambda}
\left(
\partial_{\beta} g_{\lambda\gamma}
+
\partial_{\gamma} g_{\lambda\beta}
-
\partial_{\lambda} g_{\beta\gamma}
\right).
\end{eqnarray}

The torsion tensor is given by the antisymmetric part of the affine connection,
\begin{eqnarray}
T^{\alpha}{}_{\beta\gamma}
=
\Gamma^{\alpha}{}_{\beta\gamma}
-
\Gamma^{\alpha}{}_{\gamma\beta},
\end{eqnarray}
while the nonmetricity tensor measures the variation of the metric under parallel transport,
\begin{eqnarray}
Q_{\alpha\beta\gamma}
=
\nabla_{\alpha} g_{\beta\gamma}.
\end{eqnarray}
The disformation tensor, which encodes the contribution from nonmetricity, is defined as
\begin{eqnarray}
L^{\alpha}{}_{\beta\gamma}
=
\frac{1}{2}
\left(
Q^{\alpha}{}_{\beta\gamma}
-
Q_{\beta}{}^{\alpha}{}_{\gamma}
-
Q_{\gamma}{}^{\alpha}{}_{\beta}
\right).
\end{eqnarray}

The gravitational dynamics are governed by the nonmetricity scalar $Q$, constructed as
\begin{eqnarray}
Q = Q_{\alpha\beta\gamma} P^{\alpha\beta\gamma},
\end{eqnarray}
where $P^{\alpha}{}_{\beta\gamma}$ is the nonmetricity conjugate (or superpotential), given by
\begin{eqnarray}
P^{\alpha}{}_{\beta\gamma}
&=&
-\frac{1}{4} Q^{\alpha}{}_{\beta\gamma}
+\frac{1}{4}\left(
Q_{\beta}{}^{\alpha}{}_{\gamma}
+
Q_{\gamma}{}^{\alpha}{}_{\beta}
\right)
+\frac{1}{4} Q^{\alpha} g_{\beta\gamma}\nonumber\\
&&
-\frac{1}{8}
\left(
2\tilde{Q}^{\alpha} g_{\beta\gamma}
+\delta^{\alpha}_{\beta} Q_{\gamma}
+\delta^{\alpha}_{\gamma} Q_{\beta}
\right).
\end{eqnarray}

In symmetric teleparallel gravity, both curvature and torsion are constrained to vanish, and the gravitational interaction is entirely encoded in the nonmetricity scalar $Q$. When the gravitational action is constructed linearly from $Q$, the resulting theory corresponds to the symmetric teleparallel equivalent of general relativity (STEGR), which is dynamically equivalent to Einstein's theory. Consequently, STEGR reproduces the same phenomenology as general relativity and does not address the outstanding problems associated with dark matter and dark energy.

A natural generalization is obtained by promoting $Q$ to an arbitrary function, leading to the $f(Q)$ gravity framework, analogous to the extension from GR to $f(R)$ gravity. The action of covariant $f(Q)$ gravity is given by~\cite{BeltranJimenez:2017tkd,Alwan:2024lng,Zhao:2021zab}
\begin{eqnarray}
S =
\int
\left[
\frac{1}{2\kappa} f(Q)
+
\mathcal{L}_m
\right]
\sqrt{-g}\, d^4x,
\end{eqnarray}
where $\kappa = 8\pi G/c^4$, $g$ denotes the determinant of the metric tensor, and $\mathcal{L}_m$ is the matter Lagrangian density.

Variation of this action with respect to the metric yields the field equations
\begin{eqnarray}
\frac{2}{\sqrt{-g}}
\nabla_{\alpha}\!\left(\sqrt{-g}\,f_Q\,P^{\alpha}{}_{\beta\gamma}\right)
-\frac{1}{2}g_{\beta\gamma}f\nonumber\\
+f_Q\!\left(
P_{\beta\alpha\nu}Q_{\gamma}{}^{\alpha\nu}
-2Q_{\alpha\nu\beta}P^{\alpha\nu}{}_{\gamma}
\right)
=\kappa T_{\beta\gamma},
\label{eq:12}
\end{eqnarray}
which can be recast in the covariant form~\cite{Lin:2021uqa,Zhao:2021zab,Alwan:2024lng}
\begin{eqnarray}
f_Q\,\overset{\circ}{G}_{\beta\gamma}
+\frac{1}{2}g_{\beta\gamma}\left(Qf_Q-f\right)
+2f_{QQ}\,P^{\alpha}{}_{\beta\gamma}\,\overset{\circ}{\nabla}_{\alpha}Q
=\kappa T_{\beta\gamma}.
\label{field_eq_cov_fQ}
\end{eqnarray}
Here $f_Q \equiv \partial f/\partial Q$ and $f_{QQ} \equiv \partial^2 f/\partial Q^2$, while $\overset{\circ}{G}_{\beta\gamma}$ denotes the Einstein tensor constructed from the Levi--Civita connection. In the limit $f(Q)=Q$, these equations reduce to the Einstein field equations of general relativity.

Variation with respect to the affine connection yields the equation governing the nonmetricity sector,
\begin{eqnarray}
\nabla_{\beta}\nabla_{\gamma}
\left(\sqrt{-g}\,f_Q\,P^{\beta\gamma}{}_{\alpha}\right)=0.
\label{connection_eq_cov_fQ}
\end{eqnarray}

Although it is always possible to adopt the coincident gauge, in which the affine connection vanishes identically, $\Gamma^{\alpha}{}_{\beta\gamma}=0$, the present work is formulated in the fully covariant approach where the connection is retained explicitly~\cite{Alwan:2024lng,Lin:2021uqa}.
The detailed derivation of the energy--momentum tensor in covariant $f(Q)$ gravity has been discussed extensively in the context of neutron stars~\cite{Alwan:2024lng}. Since the same formalism applies here, we do not repeat it.

In this work, we employ the above field equations to investigate the internal structure of white dwarf stars. Assuming a static and spherically symmetric spacetime and modeling the stellar matter as a perfect fluid, we derive the modified Tolman--Oppenheimer--Volkoff equations governing hydrostatic equilibrium and solve them numerically to obtain the stellar profiles and mass--radius relations.
\section{Stellar structure equations in covariant $f(Q)$ gravity}
\label{sec:3}

To describe equilibrium white dwarf configurations in covariant $f(Q)$ gravity, we consider a static and spherically symmetric spacetime. In the present work, we adopt a non-coincident gauge, i.e., $\Gamma^{\alpha}{}_{\beta\gamma}\neq0$~\cite{Alwan:2024lng,Lin:2021uqa}, so that the affine connection is treated as an independent geometrical quantity. The interior spacetime is described by the line element
\begin{eqnarray}
ds^{2}=-e^{A(r)}dt^{2}+e^{B(r)}dr^{2}
+r^{2}\left(d\vartheta^{2}+\sin^{2}\vartheta\,d\varphi^{2}\right),
\label{metric_spherical}
\end{eqnarray}
where $A(r)$ and $B(r)$ are metric functions of the radial coordinate $r$ only.

For the above metric, the non-vanishing geometrical connection components relevant to the spherically symmetric configuration are
\begin{eqnarray}
\Gamma^{\vartheta}_{r\vartheta} &= \Gamma^{\vartheta}_{\vartheta r}
= \Gamma^{\varphi}_{r\varphi}
= \Gamma^{\varphi}_{\varphi r} = \frac{1}{r}, \\
\Gamma^{r}_{\vartheta\vartheta} &= -r, \\
\Gamma^{\varphi}_{\vartheta\varphi} &= \Gamma^{\varphi}_{\varphi\vartheta}
= \cot\vartheta, \\
\Gamma^{r}_{\varphi\varphi} &= -r\sin^{2}\vartheta, \\
\Gamma^{\vartheta}_{\varphi\varphi} &= -\sin\vartheta\cos\vartheta .
\end{eqnarray}

The matter inside the star is modeled as a perfect fluid with energy--momentum tensor
\begin{eqnarray}
T_{\beta\gamma}
=
\mathrm{diag}\left(-\epsilon,\,p,\,p,\,p\right),
\end{eqnarray}
where $p$ is the fluid pressure and $\epsilon=\rho c^{2}$ is the energy density, with $\rho$ denoting the mass density.

Substituting the metric ansatz into the covariant $f(Q)$ field equations, the independent components can be written as~\cite{Lin:2021uqa,Zhao:2021zab,Alwan:2024lng}
\begin{eqnarray}
\label{eq:tt}
\kappa T_{tt} & = &
\frac{e^{A-B}}{2r^2}
\Big[
r^2 e^B f
+2 f'_Q r (e^B-1)\nonumber \\
&&+ f_Q (e^B-1)(2+rA')
+(1+e^B) r B'
\Big],
\end{eqnarray}
\begin{eqnarray}
\label{eq:rr}
\kappa T_{rr} & = &
-\frac{1}{2r^2}
\Big[
r^2 e^B f
+2 f'_Q r (e^B-1)\nonumber \\
&&+ f_Q (e^B-1)(2+rA'+rB')
-2rA'
\Big],
\end{eqnarray}
\begin{eqnarray}
\label{eq:thetatheta}
\kappa T_{\vartheta\vartheta} & = &
-\frac{r}{4e^B}
\Big[
f_Q(-4A' - rA'^2 -2rA'' + rA'B')\nonumber \\
&&+2e^B(A'+B')
+2e^B r f
-2f'_Q rA'
\Big].
\end{eqnarray}

The corresponding nonmetricity scalar is given by
\begin{eqnarray}
Q = \frac{(e^{-B}-1)(A'+B')}{r},
\end{eqnarray}
while the radial derivative of $f_Q$ is
\begin{eqnarray}
f'_Q = f_{QQ}\,\frac{dQ}{dr}.
\end{eqnarray}

To determine hydrostatic equilibrium, the gravitational field equations are supplemented by the conservation law
\begin{eqnarray}
\nabla_{\beta}T^{\beta\gamma}=0.
\end{eqnarray}
For a static and spherically symmetric perfect fluid, this gives
\begin{eqnarray}
p' = -\frac{1}{2}(\epsilon+p)A'.
\label{hydro_eq}
\end{eqnarray}
Eqs.~(\ref{eq:tt})-(\ref{eq:thetatheta}) can be recast into a system of coupled differential equations describing hydrostatic equilibrium in covariant $f(Q)$ gravity. This system yields the modified Tolman--Oppenheimer--Volkoff equations governing the internal structure of white dwarfs, together with the continuity equation following from the conservation of the energy--momentum tensor $T_{\beta\gamma}$ can be written as,
\begin{eqnarray}
A'' &=&
\frac{
2e^B \left[r\left(f(Q)+2\kappa p\right)
+f_Q(A'+B')\right]}{2f_Q r}
-\nonumber\\
&&\frac{A' \left[f_Q(4+rA'-rB')
+2f_{QQ}rQ'\right]
}{
2f_Q r
},
\label{eq:A}
\end{eqnarray}
and
\begin{eqnarray}
B' &=&
\frac{-\kappa e^B (\epsilon+p)r + f_Q A'}{f_Q},
\label{eq:B}
\end{eqnarray}
together with the hydrostatic equilibrium equation
\begin{eqnarray}
p' = -\frac{1}{2}(\epsilon+p)A'.
\label{eq:p}
\end{eqnarray}
Equations~(\ref{eq:A})--(\ref{eq:p}) constitute the modified Tolman--Oppenheimer--Volkoff system in covariant $f(Q)$ gravity. In the limit $f(Q)=Q$, they reduce to the standard Tolman--Oppenheimer--Volkoff equations of general relativity.
Regularity at the stellar center requires
\begin{eqnarray}
B(0)=0,\qquad A'(0)=0,\qquad p(0)=p_c,
\end{eqnarray}
where $p_c$ is the central pressure. The stellar surface is determined by the condition
\begin{eqnarray}
p(R)=0,
\end{eqnarray}
which defines the radius $R$.

In the exterior region, where $T_{\beta\gamma}=0$, the field equations imply
\begin{eqnarray}
A'(r)+B'(r)=0.
\end{eqnarray}
In the general relativistic limit, this leads to the Schwarzschild solution
\begin{eqnarray}
e^{A(r)}=e^{-B(r)}=1-\frac{2GM}{rc^2},
\end{eqnarray}
where $M$ is the total gravitational mass of the star.

The enclosed mass function is defined as
\begin{eqnarray}
m(r)=4\pi\int_0^r \rho(\tilde r)\,\tilde r^2\,d\tilde r,
\end{eqnarray}
and the total mass is obtained from
\begin{eqnarray}
M=m(R).
\end{eqnarray}
\section{Equation of State}
\label{sec:4}
In order to determine the equilibrium structure of white dwarfs, the stellar structure equations derived in the previous section must be supplemented by an equation of state (EoS) relating the pressure and energy density of the stellar matter. White dwarfs are supported against gravitational collapse primarily by the pressure of a completely degenerate electron gas. In this work we adopt the standard Chandrasekhar equation of state for a cold relativistic degenerate electron gas~\cite{shapiro83,Kippenhahn:2012qhp}.

Let $p_F$ denote the electron Fermi momentum. It is convenient to introduce the dimensionless relativity parameter
\begin{equation}
x=\frac{p_F}{m_e c},
\end{equation}
where $m_e$ is the electron mass and $c$ is the speed of light. The electron number density is related to the Fermi momentum through
\begin{equation}
n_e=\frac{p_F^3}{3\pi^2\hbar^3}.
\end{equation}
The mass density of the stellar matter is related to the electron number density by
\begin{equation}
\rho=\mu_e m_u n_e,
\end{equation}
where $m_u$ is the atomic mass unit and $\mu_e$ denotes the mean molecular weight per electron. For typical white dwarf compositions such as helium, carbon, or oxygen, the value $\mu_e\simeq2$ is commonly adopted~\cite{shapiro83}.

Using the relativistic Fermi gas model, the pressure of the degenerate electron gas is given by
\begin{equation}
P=\frac{\pi m_e^4 c^5}{3h^3}
\left[
x(2x^2-3)\sqrt{x^2+1}
+3\sinh^{-1}(x)
\right].
\end{equation}

Similarly, the energy density of the electron gas is
\begin{equation}
\epsilon=\frac{\pi m_e^4 c^5}{h^3}
\left[
x(1+2x^2)\sqrt{x^2+1}
-\sinh^{-1}(x)
\right].
\end{equation}

In the non-relativistic limit $(x\ll1)$ the pressure reduces to the polytropic form
\begin{equation}
P\propto\rho^{5/3},
\end{equation}
while in the ultra-relativistic regime $(x\gg1)$ it approaches
\begin{equation}
P\propto\rho^{4/3}.
\end{equation}
which corresponds to the well-known Chandrasekhar limits
for degenerate matter~\cite{Chandrasekhar1931,chandra35,shapiro83}.
The above relations provide the equation of state linking the pressure and density inside the white dwarf. In the numerical calculations, the stellar structure equations are integrated for different values of the central density $\rho_c$ using this equation of state. The integration starts from the stellar center and proceeds outward until the pressure vanishes, $P(R)=0$, which defines the radius of the white dwarf. The corresponding gravitational mass is obtained from the mass function evaluated at the stellar surface $M=m(r=R)$.
\section{Results and Discussion}
\label{sec:5}
In this section we present the numerical solutions of the modified stellar structure equations in the framework of $f(Q)$ gravity. The equations derived in the previous sections are solved together with the Chandrasekhar equation of state describing a completely degenerate electron gas. The numerical integration is performed for a range of central densities $\rho_c$ and for different positive values of the parameter $\alpha$, which characterizes the deviation from general relativity. The case $\alpha=0$ corresponds to the general relativistic limit. 
For the numerical analysis, we consider the parameter values $\alpha = 0$ and $\alpha$ ranging from $1\times10^{15}\,\mathrm{cm^{2}}$ to $1\times10^{23}\,\mathrm{cm^{2}}$. At the beginning of the numerical analysis, we note that the figures display only representative values of $\alpha$ in the range $10^{15}$–$10^{17}\,\mathrm{cm^{2}}$. For larger values of $\alpha$, the deviations from the general relativistic case become significantly larger, which would lead to very large separations between the curves and make the figures difficult to interpret. Therefore, to clearly illustrate the behavior of the stellar profiles, only a limited range of $\alpha$ is shown in the plots, while the numerical results for the wider range of $\alpha$ up to $10^{23}\,\mathrm{cm^{2}}$ are summarized in Table~\ref{table:1}.
\begin{figure}[t]
    \centering
    \includegraphics[width=0.9\linewidth]{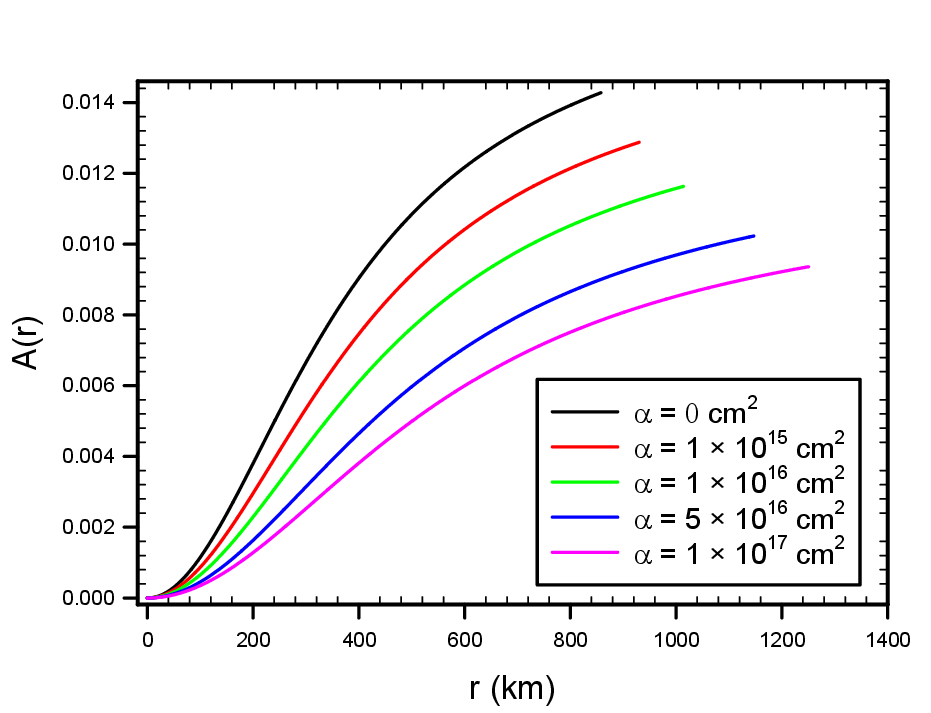}
    \caption{Radial profile of the metric function $A(r)$ inside the white dwarf for different positive values of the parameter $\alpha$ in $f(Q)$ gravity. The function increases smoothly from the stellar center toward the surface, with larger values of $\alpha$ corresponding to more extended stellar configurations compared to the general relativistic case ($\alpha=0$).}     
    \label{fig:1}
\end{figure} 
\begin{figure}[t]
    \centering
    \includegraphics[width=0.9\linewidth]{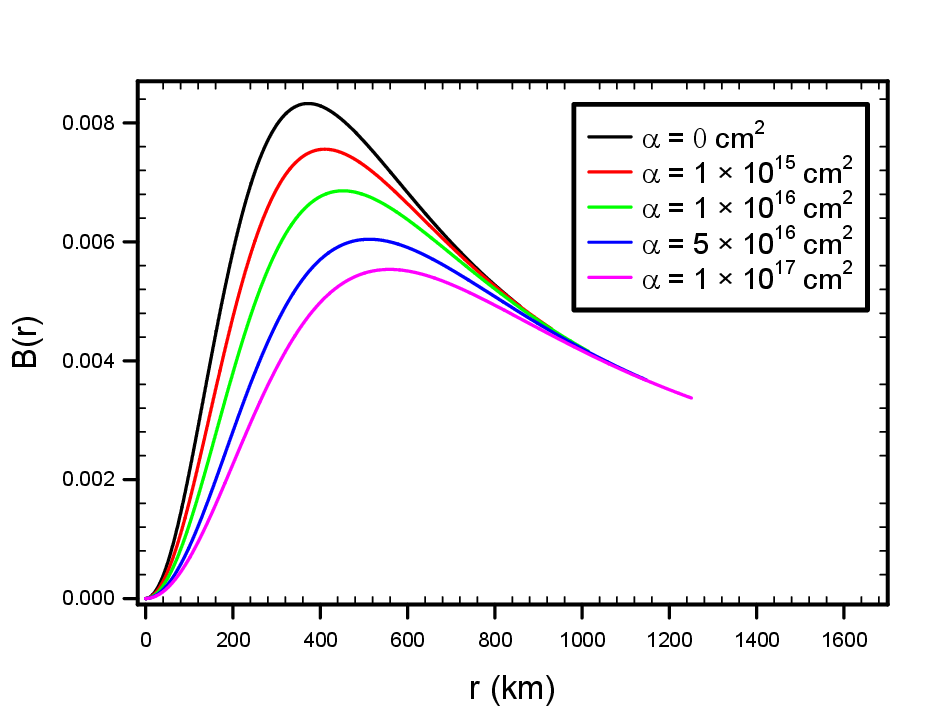}
    \caption{Radial profile of the metric function $B(r)$ inside the white dwarf for different positive values of the parameter $\alpha$ in $f(Q)$ gravity. The function increases smoothly from the stellar center toward the surface, with larger values of $\alpha$ leading to lower values of $B(r)$ compared to the general relativistic case ($\alpha=0$).}
    \label{fig:2}
\end{figure}
We note that exploratory runs with negative values of $\alpha$ were also examined. However, these cases exhibit problematic behavior at high densities and lead to configurations that deviate significantly from physically realistic white dwarf solutions. Similar difficulties associated with negative values of the parameter $\alpha$ have also been reported in studies of neutron stars within the covariant formulation of $f(Q)$ gravity, where such choices lead to unstable or nonphysical stellar configurations~\cite{Alwan:2024lng}. For this reason, the present analysis is restricted to non–negative values of $\alpha$.
\subsection{Metric functions $A(r)$ and $B(r)$}
The radial profile of the metric function $A(r)$ for different values of the parameter $\alpha$ is shown in Fig.~\ref{fig:1}. The function characterizes the interior spacetime geometry of the white dwarf and provide insight into the influence of the modified gravity parameter on the stellar structure.

The metric function $A(r)$ remains regular throughout the stellar interior and increases smoothly from the center toward the stellar surface. For the general relativistic case ($\alpha=0$), the metric function satisfies the regularity condition $A(0)=0$ at the stellar center and increases smoothly with the radial coordinate. Near the stellar surface the function reaches values of order $A(r)\sim10^{-2}$. As the parameter $\alpha$ increases, the growth of $A(r)$ becomes slightly slower, leading to lower values of $A(r)$ at a given radius. This behavior indicates that the quadratic non-metricity correction modifies the interior gravitational potential and leads to more extended stellar configurations compared to the general relativistic case. Although the qualitative behavior of the profile remains similar to the GR case, the curves gradually shift downward with increasing $\alpha$, indicating that the quadratic non-metricity correction modifies the temporal component of the spacetime metric.

The radial behavior of the metric function $B(r)$ is also shown in Fig.~\ref{fig:2}. This function remains finite at the stellar center and increases monotonically with radius. 
In the general relativistic limit ($\alpha=0$), the metric function satisfies the regularity condition $B(0)=0$ at the stellar center. As the radial coordinate increases, $B(r)$ rises to a maximum of order $10^{-3}$--$10^{-2}$ and then decreases gradually toward the stellar surface. For positive values of $\alpha$, the peak value of $B(r)$ becomes progressively smaller, indicating that the quadratic nonmetricity correction modifies the radial metric potential and leads to more extended white dwarf configurations relative to the general relativistic case.
Overall, both metric functions remain smooth and free of singularities throughout the stellar interior for all considered values of $\alpha$. The gradual change in their profiles demonstrates that the quadratic non-metricity term modifies the interior geometry of the white dwarf while preserving physically consistent stellar configurations. 
\begin{figure}[t]
    \centering
    \includegraphics[width=0.9\linewidth]{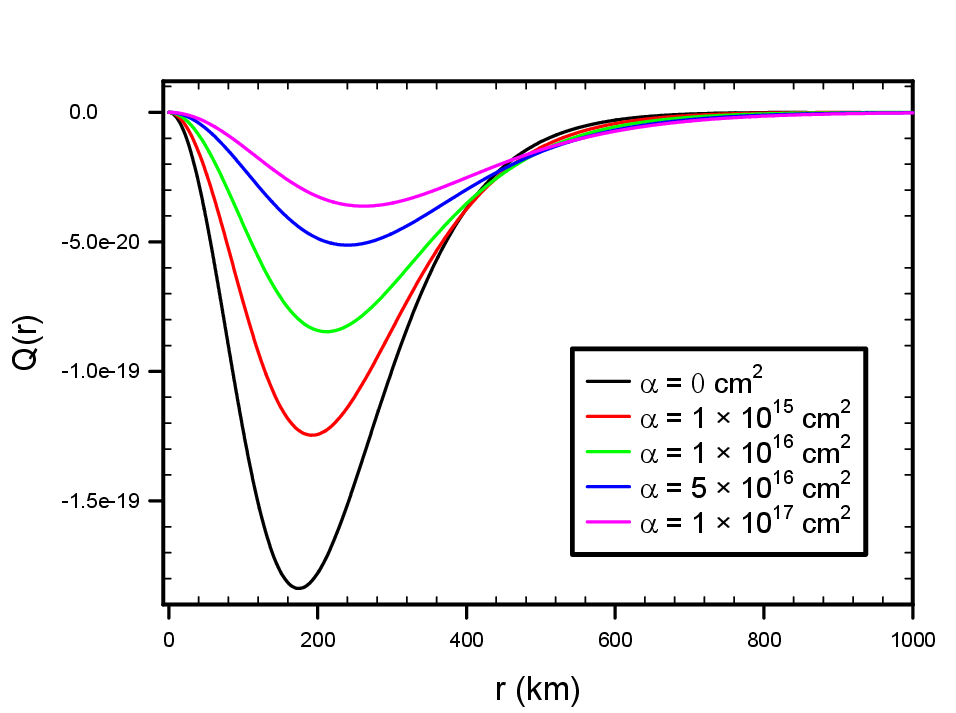}
    \caption{Radial profile of the nonmetricity scalar $Q(r)$ inside the white dwarf for different positive values of the parameter $\alpha$ in $f(Q)$ gravity. The scalar $Q(r)$ is nearly zero at the center, reaches a negative minimum in the interior, and approaches zero toward the stellar surface. Increasing $\alpha$ reduces the magnitude of this minimum compared to the general relativistic case ($\alpha=0$).}
    \label{fig:3}
\end{figure}
\subsection{Non-metricity scalar $Q(r)$}
The radial behavior of the non-metricity scalar $Q(r)$ inside the white dwarf is shown in Fig.~\ref{fig:3} for different values of the parameter $\alpha$. The scalar $Q$ plays a central role in the symmetric teleparallel formulation of gravity and characterizes the deviation from the Levi–Civita connection through the non-metricity of spacetime.
From the figure, it can be seen that the non-metricity scalar attains its maximum magnitude at the stellar center and decreases smoothly toward the surface of the star.
For the general relativistic limit ($\alpha=0$), the nonmetricity scalar satisfies $Q(0)\approx0$ at the stellar center. As the radial coordinate increases, $Q(r)$ becomes negative and reaches a minimum value of order $10^{-19}\,\mathrm{cm^{-2}}$ in the interior of the star. Moving further outward, the magnitude of $Q(r)$ gradually decreases and approaches values close to zero near the stellar surface. 
As the parameter $\alpha$ increases, the magnitude of this negative minimum decreases, indicating that the quadratic nonmetricity correction suppresses the interior nonmetricity profile. Consequently, the curves become progressively shallower for larger values of $\alpha$. The largest deviations from the general relativistic case occur in the intermediate radial region where $|Q(r)|$ attains its maximum value.
The radial profiles remain smooth and monotonic throughout the stellar interior. The magnitude of $Q(r)$ gradually decreases with increasing radial coordinate and approaches values close to zero near the stellar surface. The progressive reduction of $Q(r)$ with increasing $\alpha$ indicates that the quadratic correction in the $f(Q)=Q+\alpha Q^2$ model modifies the geometric structure of spacetime inside the star.
Overall, the results demonstrate that the non-metricity scalar remains regular throughout the stellar interior for all considered values of $\alpha$. Although the magnitude of $Q(r)$ changes with the modified gravity parameter, the solutions remain physically consistent and free of singular behavior, confirming the viability of the obtained white dwarf configurations within the explored parameter range.
\begin{figure}[t]
    \centering
    \includegraphics[width=0.9\linewidth]{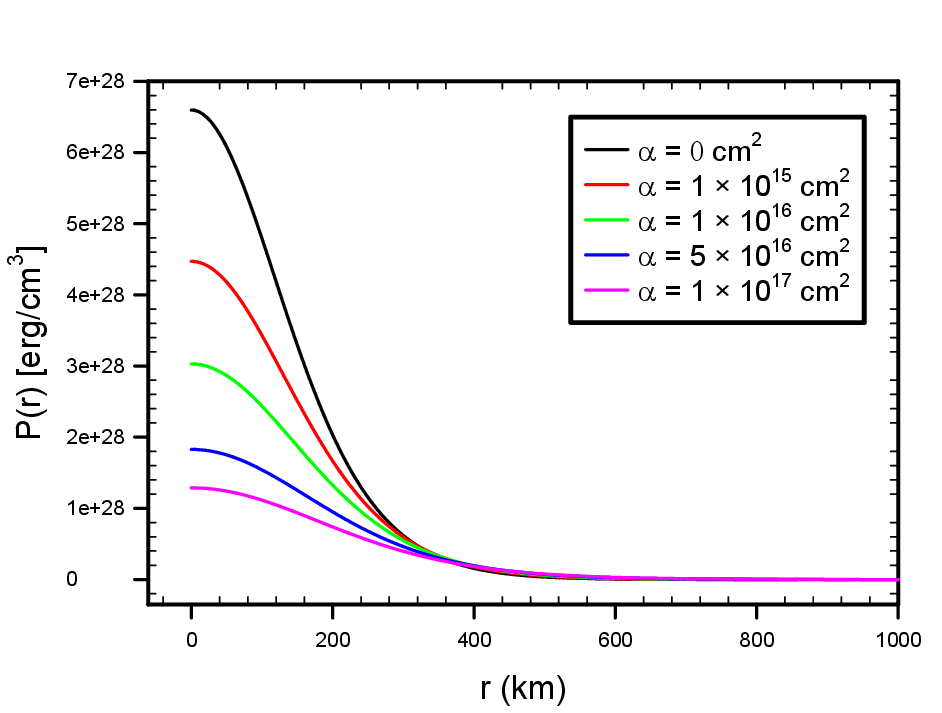}
    \caption{Radial profile of the pressure $P(r)$ inside the white dwarf for different positive values of the parameter $\alpha$ in $f(Q)$ gravity. The pressure decreases monotonically from its central value to zero at the stellar surface, while increasing $\alpha$ leads to more extended configurations compared to the general relativistic case ($\alpha=0$).}
    \label{fig:4}
\end{figure}
\subsection{Radial profiles of stellar quantities $P(r)$, $\rho(r)$ and $m(r)$}

\begin{figure}[t]
    \centering
    \includegraphics[width=0.9\linewidth]{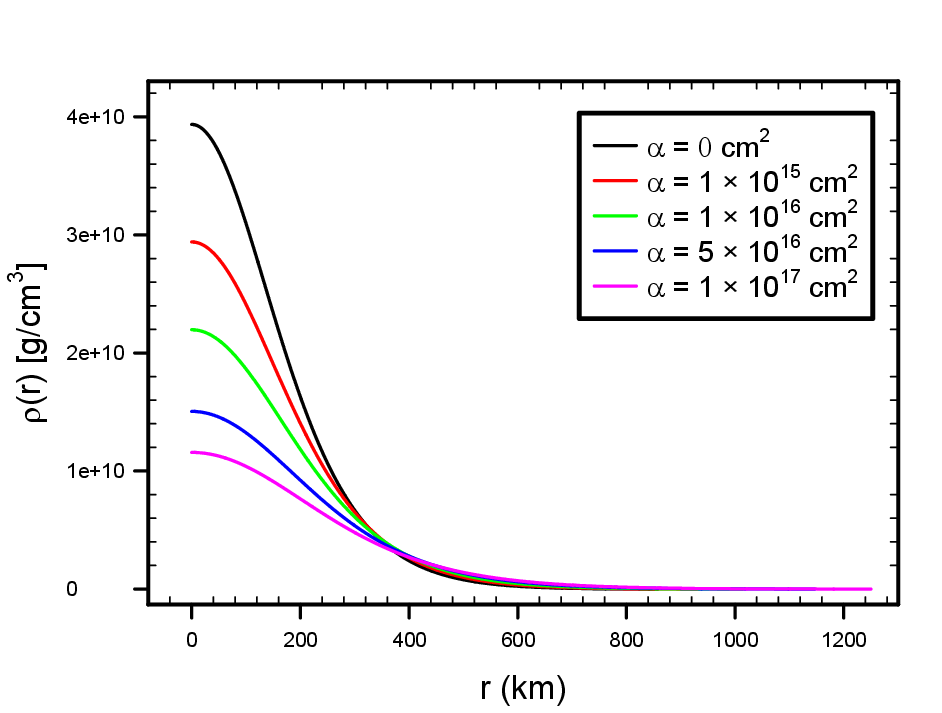}
     \caption{Radial profile of the density $\rho(r)$ inside the white dwarf for different positive values of the parameter $\alpha$ in $f(Q)$ gravity. The density decreases monotonically from the central value toward the stellar surface, while increasing $\alpha$ leads to more extended configurations compared to the general relativistic case ($\alpha=0$).}
    \label{fig:5}
\end{figure}

\begin{figure}[t]
    \centering
    \includegraphics[width=0.9\linewidth]{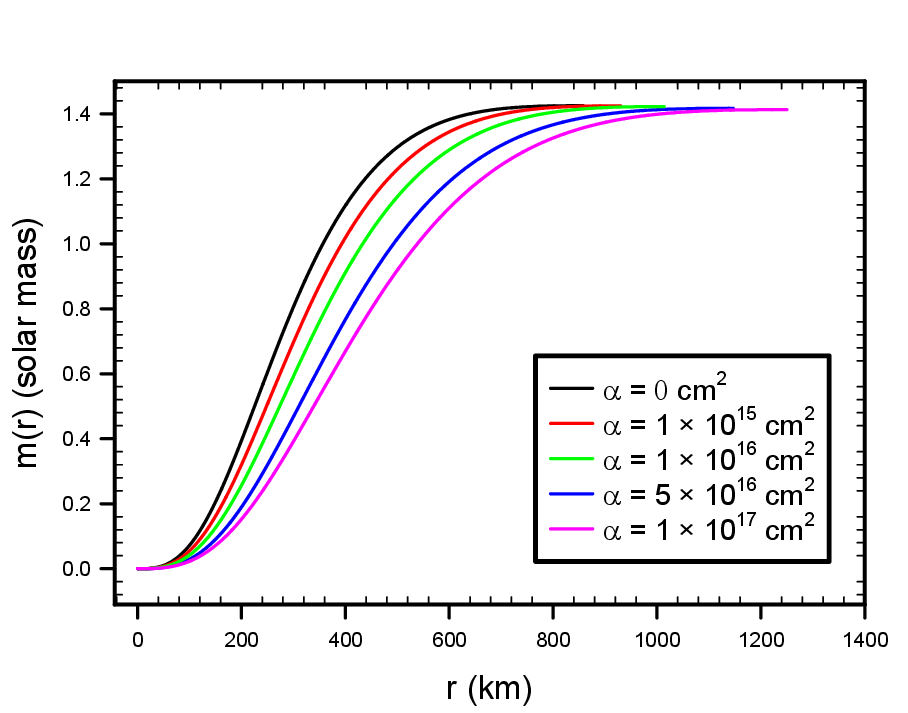}
    \caption{Radial profile of the enclosed mass $m(r)$ inside the white dwarf for different positive values of the parameter $\alpha$ in $f(Q)$ gravity. Increasing $\alpha$ modifies the mass growth and leads to larger stellar radii compared to the general relativistic case ($\alpha=0$).}
    \label{fig:6}
\end{figure}

The radial profiles of the pressure $P(r)$, density $\rho(r)$, and enclosed mass $m(r)$ for different values of the parameter $\alpha$ are shown in Figs.~\ref{fig:4},~\ref{fig:5} and ~\ref{fig:6}. These quantities describe the internal structure of the white dwarf and illustrate how the modified gravity parameter affects the equilibrium configuration.

The pressure profile $P(r)$ attains its maximum value at the stellar center and decreases monotonically with increasing radius, eventually vanishing at the stellar surface where the condition $P(R)=0$ defines the radius of the star.  
For the general relativistic limit ($\alpha=0$), the central pressure is of order $P_c \sim 10^{28}\,\mathrm{dyn\,cm^{-2}}$. As the parameter $\alpha$ increases, 
the central pressure decreases progressively, indicating that the quadratic nonmetricity correction for positive values of $\alpha$ reduces the pressure required to support the stellar configuration against gravitational collapse.

A similar behavior is observed for the density profile $\rho(r)$. The density reaches its maximum value at the stellar center and decreases smoothly toward the surface. 
For the general relativistic limit ($\alpha=0$), the density reaches its maximum value at the stellar center and decreases monotonically with increasing radial coordinate. As the parameter $\alpha$ increases, the central density decreases and the density profile becomes progressively flatter in the outer regions of the star. This behavior indicates that the quadratic nonmetricity correction modifies the internal matter distribution and leads to more extended stellar configurations compared to the general relativistic case.

The enclosed mass function $m(r)$ increases monotonically from the stellar center and reaches its maximum value at the stellar surface, corresponding to the total gravitational mass of the white dwarf. For the GR case ($\alpha=0$), the enclosed mass starts from $m(0)=0$ at the stellar center and increases monotonically with the radial 
coordinate until it reaches the total stellar mass at the surface where the pressure vanishes. As the parameter $\alpha$ increases, the growth of the mass function becomes slightly slower and the final mass of the configuration is reduced compared to the general relativistic case. At the same time, the stellar radius becomes larger, indicating that the quadratic nonmetricity correction 
produces more extended but less massive white dwarf configurations.

Overall, the pressure, density, and mass profiles remain smooth and physically well behaved throughout the stellar interior for all considered values of $\alpha$. The gradual changes observed in these quantities indicate that the quadratic non-metricity correction modifies the internal structure of the white dwarf while preserving stable stellar configurations.
\subsection{Mass--radius relation}
\begin{figure}[t]
    \centering
    \includegraphics[width=0.9\linewidth]{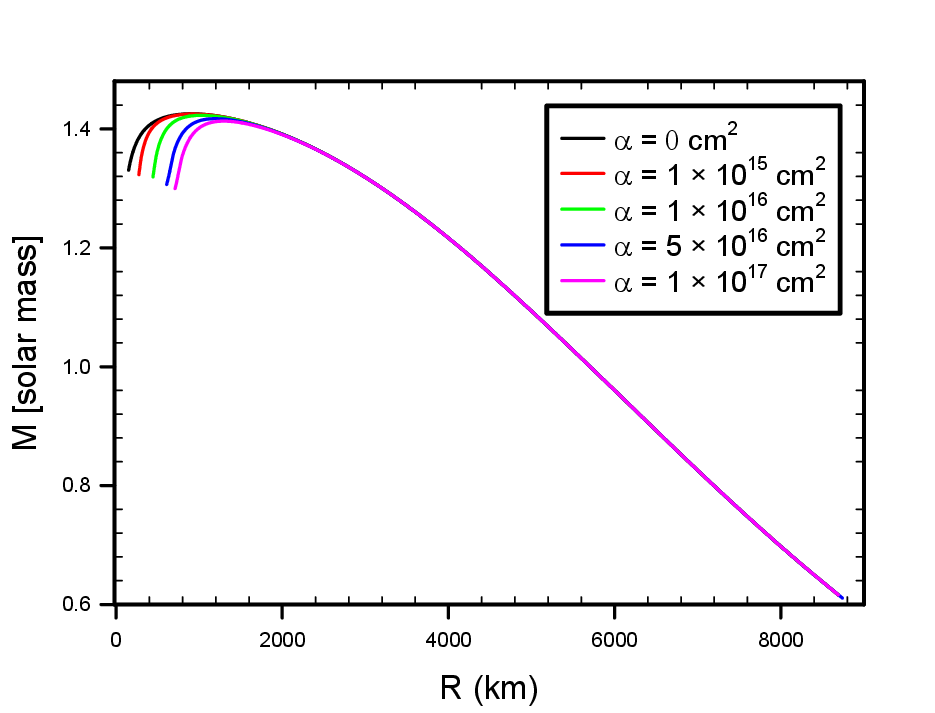} 
     \caption{Mass–radius ($M$–$R$) relation for white dwarf configurations in $f(Q)$ gravity for different positive values of the parameter $\alpha$. Increasing $\alpha$ shifts the curves toward larger radii and smaller maximum masses compared to the general relativistic case ($\alpha=0$).}
    \label{fig:7}
\end{figure}

The global equilibrium configurations of the white dwarfs are summarized through the mass--radius ($M$--$R$) relation, which is shown in Fig.~\ref{fig:7} for different values of the parameter $\alpha$. The curve corresponding to $\alpha=0$ represents the standard general relativistic limit and reproduces the familiar Chandrasekhar-type behavior.
For the GR case ($\alpha=0$), the mass increases with decreasing radius until it reaches a maximum value of approximately 
$M_{\rm max}\simeq 1.43\,M_\odot$ at a radius of about $R\simeq 1044.4671\,{\rm km}$. Beyond this point the mass decreases with decreasing radius, indicating the onset of instability in the stellar sequence.

The presence of the quadratic correction in $f(Q)$ gravity modifies the $M$--$R$ relation of white dwarfs. As the parameter $\alpha$ increases, the mass--radius curves shift toward larger radii, while the maximum mass decreases from the Chandrasekhar limit. In this analysis we consider the representative values $\alpha = 0$, $1\times10^{15}$ cm$^{2}$-$1\times10^{23}$ cm$^{2}$. These results indicate that the quadratic nonmetricity correction affects both maximum mass and the corresponding radius, producing more extended equilibrium configurations with less massive compared to the general relativistic limit. The corresponding maximum mass configurations and their associated radii for different values of $\alpha$ are listed in Table~\ref{table:1}. 
Interestingly, for $\alpha = 5\times10^{18}\,{\rm cm^2}$ the predicted mass and radius fall within the observational range of the ultra-massive white dwarf ZTF J1901+1458, which has a measured mass of $M = 1.35 \pm 0.02\,M_\odot$ and a radius of $R \approx 2140^{+160}_{-230}\,{\rm km}$~\cite{Caiazzo:2021xkk}. This agreement suggests that the quadratic $f(Q)$ correction can accommodate compact white dwarf configurations consistent with current observational constraints.
Overall, the results indicate that increasing $\alpha$ suppresses the maximum mass of the white dwarf and shifts the peak of the $M$--$R$ relation toward larger radii. This behavior reflects the influence of the quadratic non-metricity correction on the stellar equilibrium structure. Nevertheless, the qualitative form of the $M$--$R$ relation remains similar to the general relativistic case, confirming that the modified gravity model considered here still admits physically consistent white dwarf configurations.
\begin{table}[h]
\centering
\caption{Central density $\rho_c$, maximum gravitational mass $M_{max}$, and corresponding stellar radius $R$ for the white dwarf configurations obtained for different values of the parameter $\alpha$ in the quadratic $f(Q)=Q+\alpha Q^2$ gravity model. The case $\alpha=0$ corresponds to the general relativistic limit. As $\alpha$ increases, the central density required to reach the maximum mass decreases, while the stellar radius increases, leading to more extended configurations compared to GR. For $\alpha = 5\times10^{18}\,{\rm cm^2}$, the predicted mass and radius fall within the observational range of the ultra-massive white dwarf ZTF J1901+1458, which has a measured mass of $M = 1.35 \pm 0.02\,M_\odot$ and a radius of $R \approx 2140^{+160}_{-230}\,{\rm km}$~\cite{Caiazzo:2021xkk}.}

\begin{tabular}{cccc}
\toprule
$\alpha$ (cm$^2$) & $\rho_c$ (g cm$^{-3}$) & $M_{\max}$ ($M_\odot$) & $R$ (km) \\
\midrule
$0$              & $3.93\times10^{10}$ & 1.4253 & 1044.4671 \\
$1\times10^{15}$ & $2.94\times10^{10}$ & 1.4249 & 1044.5264 \\
$5\times10^{16}$ & $1.50\times10^{10}$ & 1.4172 & 1162.3499 \\
$1\times10^{17}$ & $9.71\times10^{9}$  & 1.4130 & 1327.2211 \\
$1\times10^{18}$ & $3.82\times10^{9}$ & 1.3868 & 1730.6689 \\
$5\times10^{18}$ & $1.50\times10^{9}$ & 1.3519 & 2228.8547 \\
$1\times10^{19}$ & $9.43\times10^{8}$  & 1.3304  & 2517.0015   \\
$5\times10^{19}$ & $4.68\times10^{8}$  &  1.2636 & 3014.8069   \\
$1\times10^{20}$ & $2.94\times10^{8}$  & 1.2261 & 3377.6851  \\
$5\times10^{20}$ & $1.16\times10^{8}$  & 1.1130 & 4206.9064  \\
$1\times10^{21}$ & $6.46\times10^{7}$  & 1.0513 & 4785.1959  \\
$5\times10^{21}$ & $2.69\times10^{7}$  & 0.8407 & 5768.5839  \\
$1\times10^{22}$ & $2.01\times10^{7}$  & 0.8100 & 6126.0773  \\
$1\times10^{23}$ & $4.69\times10^{6}$  & 0.5452 & 8106.3445 \\
\bottomrule
\end{tabular}
\label{table:1}
\end{table} 
\begin{figure}[t]
    \centering
    \includegraphics[width=0.9\linewidth]{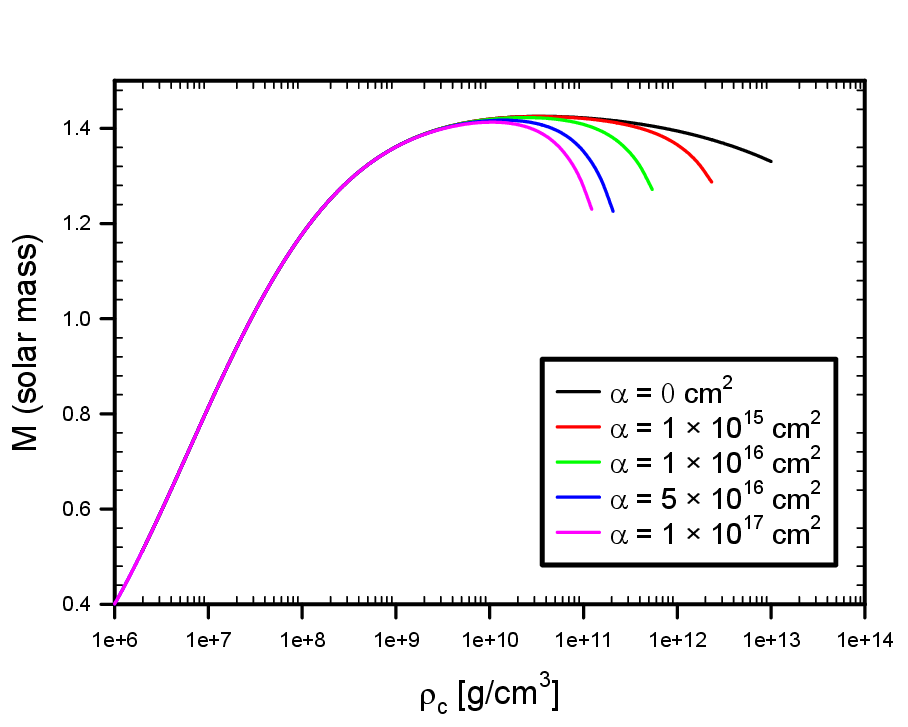}
     \caption{Stellar mass $M$ as a function of the central density $\rho_c$ for white dwarf configurations in $f(Q)$ gravity for $\alpha = 0$, $1\times10^{15}$, $1\times10^{16}$, $5\times10^{16}$, and $1\times10^{17}$ cm$^{2}$. The mass increases with $\rho_c$ up to a maximum value, beyond which the configurations become unstable. Increasing $\alpha$ shifts the peak toward lower central densities and reduces the maximum mass compared to the general relativistic case.}
    \label{fig:8}
\end{figure}
\subsection{Mass and radius as functions of the central density}
\begin{figure}[t]
    \centering
    \includegraphics[width=0.9\linewidth]{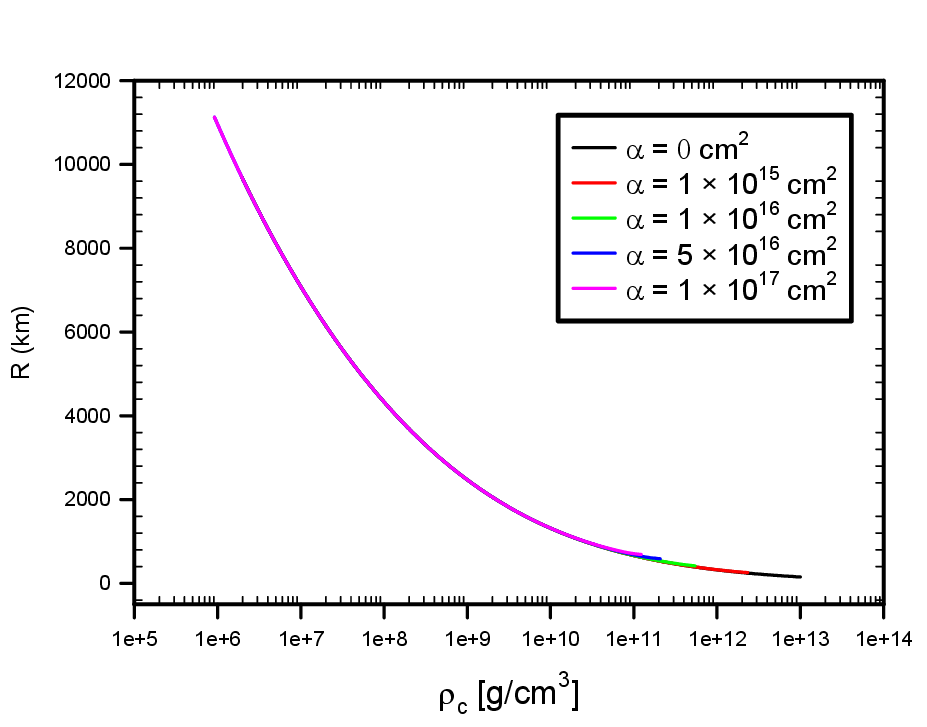}
    \caption{Stellar radius $R$ as a function of the central density $\rho_c$ for white dwarf configurations in $f(Q)$ gravity for $\alpha = 0$, $1\times10^{15}$, $1\times10^{16}$, $5\times10^{16}$, and $1\times10^{17}$ cm$^{2}$. The radius decreases monotonically with increasing central density. Increasing $\alpha$ increases the stellar radius for a given $\rho_c$, indicating less compact configurations compared to the general relativistic case.}
    \label{fig:9}
\end{figure}

The global properties of the white dwarf configurations are summarized in Figs.~\ref{fig:8} and ~\ref{fig:9}, where the gravitational mass $M$ and the corresponding stellar radius $R$ are plotted as functions of the central density $\rho_c$ for different values of the parameter $\alpha$.

The mass–central density relation $M(\rho_c)$ exhibits the characteristic behavior expected for white dwarf sequences. As the central density increases, the gravitational mass initially increases, indicating that more massive configurations can be supported by the degeneracy pressure of the electron gas. However, beyond a certain value of $\rho_c$, the mass reaches a maximum value and begins to decrease, signaling the onset of instability in the stellar configurations. The curve corresponding to $\alpha=0$ reproduces the standard behavior obtained within general relativity.
The presence of the quadratic correction in $f(Q)$ gravity modifies this relation. As the value of $\alpha$ increases, the entire $M(\rho_c)$ curve shifts, leading to noticeable changes in the maximum mass and the corresponding central density. These deviations indicate that the non-metricity correction influences the equilibrium structure and stability properties of white dwarfs.

The radial size of the star is also affected by the modified gravity parameter. The radius–central density relation $R(\rho_c)$ shows that the stellar radius decreases with increasing central density, reflecting the stronger gravitational compression in denser configurations. The inclusion of the parameter $\alpha$ leads to quantitative changes in this relation, altering the radius corresponding to a given central density.
Overall, the results demonstrate that the quadratic $f(Q)$ correction modifies both the mass and radius of white dwarf configurations while preserving the qualitative behavior expected from compact stellar objects. These deviations from the general relativistic predictions provide insight into the role of non-metricity in determining the global properties of white dwarfs.
\subsection{Stability criterion}
In order to examine the stability of the obtained stellar configurations, we analyze the behavior of the relativistic adiabatic index $\Gamma$, which provides an important criterion for dynamical stability. The adiabatic index is defined as
\begin{equation}
\Gamma = \frac{\rho c^2+ P}{c^2 P}\frac{dP}{d\rho}.
\end{equation}

For a relativistic compact object to remain stable against radial perturbations, the adiabatic index must satisfy the well-known condition
\begin{equation}
\Gamma > \frac{4}{3}.
\end{equation}

\begin{figure}[t]
    \centering
    \includegraphics[width=0.9\linewidth]{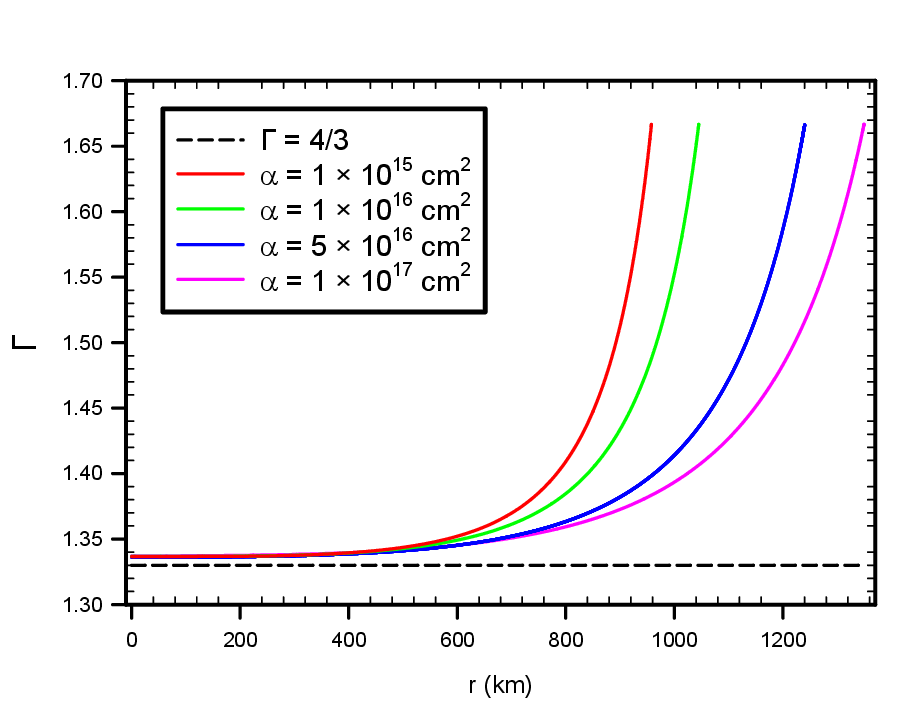}
    \caption{Radial profile of the relativistic adiabatic index $\Gamma(r)$ inside the white dwarf for different positive values of the parameter $\alpha$ in $f(Q)$ gravity. The dashed line represents the stability condition $\Gamma = 4/3$. For all considered values of $\alpha$, $\Gamma$ remains above this limit throughout the stellar interior, indicating dynamical stability against radial perturbations.}
    \label{fig:10}
\end{figure}

Figure~\ref{fig:10} shows the radial variation of the adiabatic index for different values of the parameter $\alpha$. The curves correspond to $\alpha = 0$, $10^{15}$, $10^{16}$, $5\times10^{16}$, and $10^{17}\,{\rm cm^2}$. In all cases the value of $\Gamma$ remains greater than the critical value $4/3$ throughout the stellar interior, indicating that the obtained white dwarf configurations are dynamically stable.
Although the value of $\Gamma$ decreases with increasing $\alpha$, it remains well above the critical threshold required for stability.
These results demonstrate that the inclusion of the quadratic non-metricity correction in $f(Q)$ gravity does not destabilize the white dwarf configurations within the explored parameter range. The stellar models therefore satisfy the standard dynamical stability criterion while exhibiting modified structural properties due to the presence of the parameter $\alpha$.
\section{Summary and Conclusions}
\label{sec:6}
In this work, we investigated the equilibrium structure of white dwarfs in the framework of quadratic $f(Q)$ gravity, adopting the model $f(Q)=Q+\alpha Q^{2}$,
where $\alpha$ is the coupling parameter controlling deviations from general relativity (GR). The stellar equilibrium equations were obtained from the modified Tolman--Oppenheimer--Volkoff (TOV) system in covariant $f(Q)$ gravity and solved numerically together with the Chandrasekhar equation of state for cold, completely degenerate white dwarf matter.
For each stellar configuration, we determined the radial profiles of the metric functions $A(r)$ and $B(r)$, the pressure $p(r)$, the density $\rho(r)$, the enclosed mass $m(r)$, and the nonmetricity scalar $Q(r)$. From these solutions, we constructed the mass--radius relation and examined the effect of the quadratic correction parameter $\alpha$ on the global properties of white dwarfs.

As a consistency check, we first considered the GR limit, $\alpha=0$, for which the numerical solutions reproduce the standard Chandrasekhar white dwarf sequence. In this case, the mass increases with central density up to a maximum value of approximately $M_{\max}\approx1.43\,M_\odot$, beyond which the sequence turns over, signaling the onset of instability. The corresponding radii range from $\sim10^{4}\,\mathrm{km}$ for low-mass configurations to $\sim10^{3}\,\mathrm{km}$ near the maximum-mass state.

We then examined the influence of positive values of $\alpha$. The results show that the quadratic nonmetricity correction primarily affects the high-density part of the stellar sequence. For low central densities, the solutions remain very close to the GR case, indicating that the quadratic term has only a small effect in the weak-field regime. As the central density increases, however, the deviations become more pronounced. In particular, increasing $\alpha$ reduces the maximum mass and shifts the corresponding equilibrium configurations toward larger radii. This behavior shows that the quadratic correction modifies the balance between gravity and electron degeneracy pressure, leading to less compact white dwarf configurations than in GR.
Negative values of $\alpha$ were also explored in preliminary calculations. These cases were found to generate unstable or nonphysical configurations at high densities, and were therefore not included in the detailed analysis.
We also examined the radial profile of the relativistic adiabatic index $\Gamma$. For all configurations considered in the present work, $\Gamma$ remains above the critical value $4/3$ throughout the stellar interior. This supports the conclusion that the obtained solutions satisfy the standard local stability criterion, while the turning-point behavior in the mass--central density relation provides an additional indicator of the onset of instability along the equilibrium sequence.
Overall, our results show that quadratic $f(Q)$ gravity can produce appreciable modifications to white dwarf structure in the relativistic density regime, while leaving the low-density regime close to the GR prediction. In particular, the deviations become significant near the Chandrasekhar limit, suggesting that massive white dwarfs can serve as useful astrophysical probes of nonmetricity-based modifications of gravity.

In summary, we have shown that physically acceptable white dwarf configurations can be constructed in the quadratic $f(Q)$ model and that their mass--radius relation departs systematically from the GR result as $\alpha$ increases. An especially interesting result is that, for suitable values of $\alpha$, the predicted mass and radius become compatible with the observed properties of the ultra-massive white dwarf ZTF J1901+1458. Future work may extend the present analysis by including finite-temperature effects, magnetic fields, rotation, or more realistic stellar compositions, which could further constrain the parameter space of modified gravity models.
We also analyzed the radial profile of the relativistic adiabatic index $\Gamma$ inside the stellar interior. The adiabatic index remains greater than the critical value $4/3$ throughout the star and exhibits a smooth radial variation from the center toward the stellar surface. This behavior indicates that the stellar configurations remain dynamically stable against radial perturbations. Furthermore, the profiles of $\Gamma$ remain well behaved for all considered values of the parameter $\alpha$, confirming that the obtained solutions satisfy the standard stability conditions for compact stellar objects.
The numerical results therefore indicate that quadratic $f(Q)$ gravity produces measurable modifications in the equilibrium structure of compact stars while leaving the weak-field regime essentially unchanged. In particular, the deviations from GR become significant only near the relativistic density regime close to the Chandrasekhar limit. These findings suggest that compact stars, and especially white dwarfs near their maximum mass, can provide useful probes for testing deviations from GR in the framework of symmetric teleparallel gravity.

In summary, we have demonstrated that white dwarf configurations can be consistently constructed within the quadratic $f(Q)$ gravity model and that the resulting mass–radius relation shows clear deviations from the GR prediction at high densities. Future investigations could extend the present analysis by considering finite-temperature effects, magnetic fields, or more realistic compositions of stellar matter, which may further constrain the parameter space of modified gravity theories.


\bibliographystyle{elsarticle-harv}
\bibliography{reference}


%
%


\end{document}